# Emergent spin-glass state in the doped Hund's metal $CsFe_2As_2$

S. J. Li[1,*], D. Zhao[1,*], S. Wang[2,*], S. T. Cui[2], N. Z. Wang[1], J. Li[1], D. W. Song[1], B. L. Kang[1], L. X. Zheng[1], L. P. Nie[1], Z. M. Wu[1], Y. B. Zhou[1], M. Shan[1], Z. Sun[2,3,4,5,†], T. Wu[1,3,4,5,6,†] and X. H. Chen[1,3,4,5,6,†]

[1]Hefei National Research Center for Physical Sciences at the Microscale, University of Science and Technology of China, Hefei, Anhui 230026, China

[2]National Synchrotron Radiation Laboratory, University of Science and Technology of China, Hefei, 230029, China

[3]CAS Key Laboratory of Strongly-coupled Quantum Matter Physics, Department of Physics, University of Science and Technology of China, Hefei, Anhui 230026, China

[4]Collaborative Innovation Center of Advanced Microstructures, Nanjing University, Nanjing 210093, China

[5]Hefei National Laboratory, University of Science and Technology of China, Hefei 230088, China

[6]CAS Center for Excellence in Superconducting Electronics (CENSE), Shanghai 200050, China

*these authors contributed equally to this work
†Corresponding author. Email: zsun@ustc.edu.cn (Z. Sun); wutao@ustc.edu.cn (T. Wu); chenxh@ustc.edu.cn (X.H. Chen)

## ABSTRACT

Hund's metal is one kind of correlated metal, in which the electronic correlation is strongly influenced by the Hund's interaction. At high temperatures, while the charge and orbital degrees of freedom are quenched, the spin degrees of freedom can persist in terms of frozen moments. As temperature decreases, a coherent electronic state with characteristic orbital differentiation always emerges at low temperatures through an incoherent-to-coherent crossover, which has been widely observed in iron-based superconductors (e.g., iron selenides and $AFe_2As_2$ (A = K, Rb, Cs)). Consequently, the above frozen moments are "screened" by coupling to orbital degrees of freedom, leading to an emergent Fermi-liquid state. In contrast, the coupling among frozen moments should impede the formation of the Fermi-liquid state by competitive magnetic ordering, which is still unexplored in Hund's metal. Here, in the iron-based Hund's metal CsFe$_2$As$_2$, we adopt a chemical substitution at iron sites by Cr/Co atoms to explore the competitive magnetic ordering. By a comprehensive study of resistivity, magnetic susceptibility, specific heat and nuclear magnetic resonance, we demonstrate that the Fermi-liquid state is destroyed in Cr-doped CsFe$_2$As$_2$ by a spin-freezing transition below $T_g$ ~ 22 K. Meanwhile, the evolution of charge degrees of freedom measured by angle-resolved photoemission spectroscopy also supports the competition between the Fermi-liquid state and spin-glass state.

## I. INTRODUCTION

Understanding the role of electronic correlation in high-$T_c$ superconductors is a long-standing challenge in condensed matter physics [1,2]. In high-$T_c$ cuprate superconductors, the strong electronic correlation originates from a large on-site Coulomb repulsion (U) exceeding the bandwidth (Δ) (U > Δ). The resulting Mott physics is believed to be a key to high-$T_c$ superconductivity [1]. In contrast, due to multiorbital character, the electronic correlation in iron-based superconductors (IBS) is strongly influenced by Hund's interaction ($J_H$), which leads to a new kind of correlated electronic liquid named Hund's metal [3-11]. In this case, the electronic system evolves from incoherent atomic states at high temperatures to a coherent state at low temperatures. At intermediate temperatures, charge and orbital degrees of freedom are itinerant, whereas spin degrees of freedom behave localized [2,4,7,11]. A prime feature of Hund's metal is the incoherent-to-coherent crossover [5,11], which has been observed in the heavily hole-doped iron-pnictide superconductors [12-14] [Fig. 1(c)]. In addition, similar crossover behaviors have been observed in iron-selenide superconductors as well [15,16].

As the effect of Hund's interaction increases, the Hund's metal also exhibits remarkable orbital differentiation [3,7,17-23] [Fig. 1(b)]. In iron-based superconductors, an extreme case is called the orbital-selective Mott state, in which the $d_{xy}$ orbital remains completely localized down to zero temperature [16,17,20,23]. Whether an explicit orbital-selective Mott state exists is still under debate in iron-based superconductors [24]. On the other hand, the frozen moment, which is defined by a finite spin-spin correlation function over long times [as shown in the inset of Fig. 1(d)], has been proposed to depict the spin degrees of freedom in the incoherent phase [4,11,25]. As a result, the incoherent state at intermediate temperatures is described as a non-Fermi-liquid state with frozen moments, which displays an orbital-dependent fractional power-law behavior in their quasiparticle self-energy [7,8,11,25].

In iron-pnictide BaFe$_2$As$_2$, the correlated electronic state presents a continuous evolution by carrier doping [26-29]. As electron doping increases, the electronic correlation effect is reduced, leading to a conventional Fermi-liquid state [Fig. 1(d)]. In contrast, as the hole doping increases, the electronic correlation effect is enhanced, as evidenced by the enhanced electronic effective mass and orbital differentiation [3,21,27] [Figs. 1(a) and 1(b)]. In the extreme case of AFe$_2$As$_2$ (A = K, Rb, Cs) with the $3d^{5.5}$ configuration, although the electronic correlation is largely enhanced, an emergent Fermi-liquid state eventually appears at low temperatures [12-14,30], suggesting a similar screening effect of the frozen moment as heavy fermion materials [31]. Meanwhile, inelastic neutron scattering and nuclear magnetic resonance (NMR) experiments indicated the existence of significant antiferromagnetic correlations in AFe$_2$As$_2$ (A = K, Rb, Cs) [32,33], and a possible quantum criticality has also been reported [34]. Considering the above facts, the competing interactions on frozen moments could result in competition between the Fermi-liquid state and magnetic ordering state. Phenomenally, this is similar to the Doniach phase diagram in heavy fermion materials [8,31,35] but the physics of Hund's metal is fundamentally different from that in heavy fermions. Here, in order to explore the above competition between Fermi-liquid state and magnetic ordering state, we try to dilute the coherent interaction between iron sites by chemical substitution with chromium (Cr) and cobalt (Co) atoms in CsFe$_2$As$_2$. A spin-freezing transition is observed below $T_g$ ~ 22 K in Cr-doped CsFe$_2$As$_2$, which destroys the Fermi-liquid state in pristine CsFe$_2$As$_2$. In contrast, the Fermi-liquid state is preserved in Co-doped CsFe$_2$As$_2$. Next, we will show the results of different measurements in detail mainly performed on the Cr-doped (x = 0.10), pristine and Co-doped (x = 0.11) CsFe$_2$As$_2$ respectively, unless otherwise stated (for detailed sample information, see Sec. II (Methods) section and Secs. S1 and S2 of the Supplemental Material [36]).

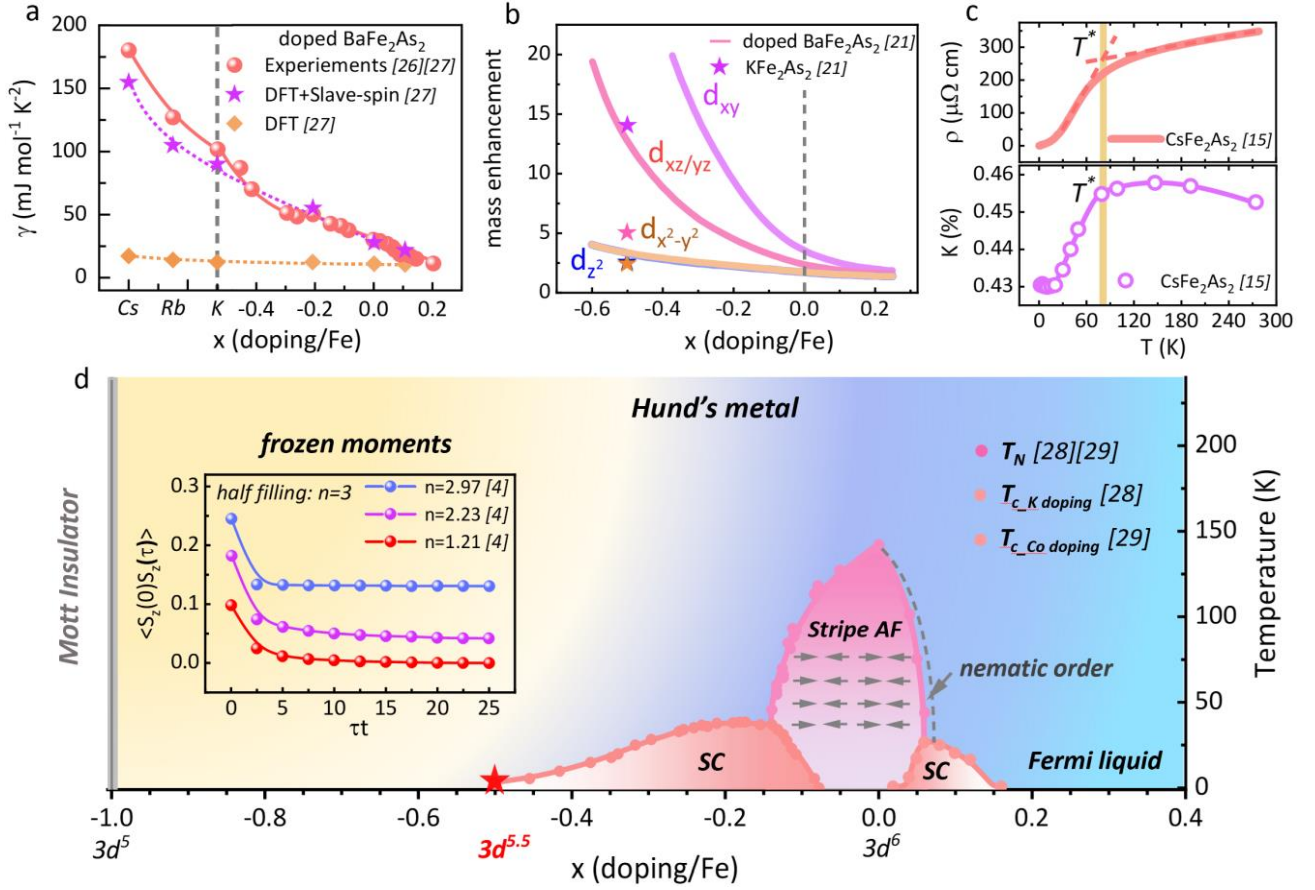

FIG. 1. Hund's metal and its doping-dependent evolution in iron-pnictide superconductors with the 122 structures. The positive value x is on behalf of the electron-doping (Co-doping) contents while the negative value x is on behalf of the hole-doping (K-doping) contents for per iron site. (a) Sommerfeld coefficient $\gamma_n$ (red symbols) of the linear-in-temperature contribution to specific heat at low temperature in the normal phase in the Ba-122 family [26,27]. Purple stars represent density-functional theory + slave-spin (DFT+SS) calculations for the tetragonal paramagnetic phase. The yellow diamonds show the results of the density functional theory (DFT) calculations for the uncorrelated electrons. Such a noteworthy discrepancy between DFT calculations and the experimental results indicates that electronic correlations play an important role in the Ba-122 family, especially in $AFe_2As_2$ (A=K, Rb, Cs). Lines are guides to the eye. (b) Orbitally resolved mass enhancement calculated within local-density approximation (LDA) + slave-spin mean-field (SS) theory for doped $BaFe_2As_2$ (solid lines) and $KFe_2As_2$ (stars) [21]. Considering the increase of covalency in the Fe-As planes introduced by the Ba-K substitution, the calculated value of mass enhancement is close to the experimental ones for $KFe_2As_2$, marked by the stars. (c) Top panel: Temperature-dependent resistivity for $CsFe_2As_2$ [13]. Dashed lines are two tentative $T$-linear trends to extract the value of $T^*$, which is the characteristic

temperature of the coherent-incoherent crossover behavior. Bottom panel: Temperature-dependent Knight shift (K) for CsFe$_2$As$_2$ [13]. Purple line is a guide for the eyes. Bold yellow lines mark nearly the same coherent-incoherent crossover temperature $T^*$ determined by both the Knight shift and resistivity. (d) Sketch of the phase diagram of iron-pnictide superconductors with the 122 structures. Experimentally measured superconducting (SC) and stripe-type antiferromagnetic (AF) phases around the $3d^6$ configuration are indicated by the pinkish-orange and light-magenta areas, respectively. $T_N$ is the AF transition temperature [28,29]. $T_{c\_K\ doping}$ is the SC transition temperature for Ba$_{1-x}$K$_x$Fe$_2$As$_2$ [28]. $T_{c\_Co\ doping}$ is the SC transition temperature for Ba(Fe$_{1-x}$Co$_x$)$_2$As$_2$ [29]. The area under the dashed gray line denotes the nematic-order phase. In the heavily electron-doping region, the Fermi-liquid phase appears. In the intermediate region between the $3d^6$ and $3d^5$ configuration, the system is described as Hund's metal in which the correlation effects are driven by Hund's interaction [4,7,9,23]. At the extreme $3d^5$ configuration, the putative Mott insulator phase would appear, which is denoted by the gray area in the left-most region. Here, our present study is focused on the study of CsFe$_2$As$_2$ and its dopants (around the $3d^{5.5}$ configuration), highlighted by the red star. (Inset) Imaginary time ($\tau$) dependence of the spin-spin correlation function $\langle S_z(0)S_z(\tau)\rangle$ with a unique choice of the interaction parameters ($U = 8t, J_H/U = 1/6$) and different carrier concentrations ($n$)[4], based on a single-site dynamical mean-field study of a three-band model relevant to the transition-metal oxides with half-filling $n = 3$. Here, $U$ is the intraorbital Coulomb interaction, $J_H$ is the coefficient of the Hund's coupling, and $t$ is the bandwidth of the semicircular density of states. Such results revealed a quantum phase transition between a paramagnetic metallic phase and an incoherent metallic phase with frozen moments. Specifically, when the carrier concentration is greater than the critical value ($n > n_c$), the spin-spin correlation function is seen to approach a constant at long times rather than decay to zero, indicating the presence of frozen moments.

## II. METHODS
### A. Sample preparation

High-quality Cr-doped and Co-doped CsFe$_2$As$_2$ single crystals were synthesized by the self-flux method. First, the Cs chunks, Fe, Cr (Co) and As powders were weighed according to the molar ratio of Cs: Fe: Cr (Co): As = 6: 1-x: x: 6. Then, the following operations were the same as those of the synthesis of the CsFe$_2$As$_2$ single crystal, as reported in Ref. [45]. Notably, the maximum temperature $T_{max}$ during the synthesis of the doped samples would vary with the nominal doping content x. Based on practical experiences, $T_{max}$ could be approximated via the formula $T_{max} = \frac{(1-x)*T_{m,Fe}+x*T_{m,Cr(Co)}}{T_{m,Fe}} * 950\ °C$, where 950 °C was the maximum temperature during the synthesis of CsFe$_2$As$_2$ and $T_{m,A}$ represented the melting temperature of different elements. Here, taking the Cr (Co)-doped CsFe$_2$As$_2$ with the nominal doping content x = 0.15 as an example, $T_{max}$ would be fixed to 980 °C (945 °C). In addition, the cooling rate could be adjusted properly to obtain samples of different sizes. Doped samples with a higher doping content (x > 0.15) or lower doping content (x < 0.05) could not be obtained at present. The actual doping content of the mainly reported Cr (Co)-doped CsFe$_2$As$_2$ in this text is x = 0.10 (x = 0.11) with the nominally doping content x = 0.15 (for detailed EDS results, see Sec. S2 of the Supplemental Material [36]).

### B. Electronic transport, magnetization and specific-heat measurements

The temperature-dependent resistivity by a standard *dc* four-probe method and the specific heat were measured on a commercial PPMS-14T (Physical Property Measurement System) (Quantum Design). The direct current (*dc*) and alternating current (*ac*) magnetic susceptibility were measured on a VSM-7T (Vibrating Sample Magnetometer) (Quantum Design). X-ray diffraction (XRD) characterization was performed on a SmartLab-9 diffractometer (Rikagu) from 5º to 70º with a scanning rate of 4º per minute. The actual doping contents of different doped samples were determined by EDS measurements, mounted on the field emission scanning electronic microscope, Sirion 200.

### C. NMR measurements

Standard NMR spin-echo techniques were used with a commercial NMR spectrometer from Thamway Co. Ltd. The external magnetic field was generated by a 12-T magnet from Oxford Instruments. The samples were compactly packed into the copper coils, which were simultaneously used to calibrate the external field from the $^{63}$Cu NMR signal. The $^{75}$As and $^{133}$Cs NMR spectra were obtained by sweeping the frequency at a fixed magnetic field of 12 T and then integrating the spin-echo signal over the full frequency range at various temperatures. Spin-lattice relaxation time ($T_1$) was extracted by fitting the spin-echo decay curve with the formula $M_t = M_0 + M_1(0.1e^{-\left(\frac{t}{T_1}\right)^r} + 0.9e^{-\left(\frac{6t}{T_1}\right)^r})$ for $^{75}$As nuclei (nuclear spin $I = 3/2$) in Co-doped CsFe$_2$As$_2$ and $M_t = M_0 + M_1 e^{-\left(\frac{t}{T_1}\right)^r}$ for $^{133}$Cs nuclei (nuclear spin $I = 7/2$) in Cr-doped CsFe$_2$As$_2$. Here, the relaxation quantity $M$ is the intensity of spin echo, $t$ is time, $T_1$ is the spin-lattice relaxation time, and $r$ is the stretching exponent

($0 < r \leq 1$).

**D. Angle-resolved photoemission spectroscopy measurements**

Angle-resolved photoemission spectroscopy (ARPES) experiments were performed at the beamline 13U of the National Synchrotron Radiation Laboratory at Heifei, China, equipped with a Scienta Omicron DA30L analyzer. The angular resolution was 0.3°, and the combined instrumental energy resolution was better than 15 meV. All samples were cleaved and measured under a vacuum better than $1 \times 10^{-10}$ mbar. During measurements, the spectroscopy qualities were carefully monitored to avoid the sample aging issue.

## III. RESULTS

### A. Spin freezing induced a non-Fermi-liquid state

As shown in Fig. 2(a), the temperature-dependent in-plane resistivity ($\rho_{ab}$) of Co-doped CsFe$_2$As$_2$ retains metallic behavior down to 2 K, similar to that in pristine CsFe$_2$As$_2$. In contrast, the temperature-dependent $\rho_{ab}$ of Cr-doped CsFe$_2$As$_2$ displays an insulating-like behavior below 100 K, suggesting a quite different charge ground state induced by Cr doping. Notably, the value of $\rho_{ab}$ in Cr-doped CsFe$_2$As$_2$ is increased by three orders of magnitude compared to that of the pristine CsFe$_2$As$_2$ at 2 K. To further clarify the nature of spin degrees of freedom, we also measured the temperature-dependent in-plane *dc* magnetic susceptibility ($\chi_{dc}$) for both Co-doped and Cr-doped CsFe$_2$As$_2$. As shown in Fig. 2(b), while $\chi_{dc}$ of Co-doped CsFe$_2$As$_2$ exhibits similar paramagnetic behavior as that in pristine CsFe$_2$As$_2$, $\chi_{dc}$ of Cr-doped CsFe$_2$As$_2$ displays a clear peaklike behavior around $T_g \sim 22$ K, and an apparent bifurcation between the zero-field cooling (ZFC) and field-cooling (FC) curves happens just below $T_g$, suggesting a possible spin-freezing behavior induced by Cr doping. Similar behaviors in resistivity and magnetic susceptibility have already appeared in Cr-doped CsFe$_2$As$_2$ with x = 0.06, which is the sample with the least doping content at the present stage (see Sec. S3 of the Supplemental Material [36]). Here, we mention that the spin-freezing/glass behavior usually comes from frustrated magnetic interactions. The freezing temperature can be dependent on many factors. In our case, there are at least two factors important for determining the freezing temperature: the strength of magnetic interaction and the doping level of Cr. The same value of $T_g$ in the two different doping level might be an accident caused by the change of the above two factors.

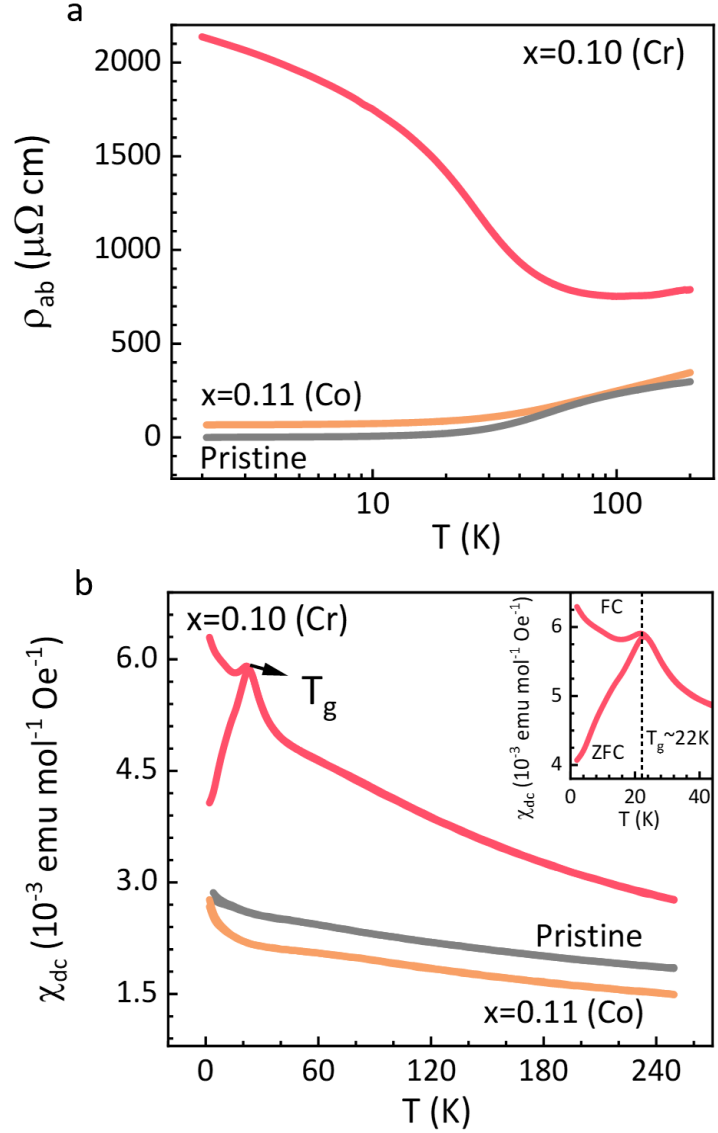

FIG. 2. Electrical transport and *dc* magnetic susceptibility evidence for spin freezing and non-Fermi-liquid behavior in Co/Cr-doped CsFe$_2$As$_2$. (a) Temperature-dependent in-plane resistivity $\rho_{ab}$ of pristine and Co/Cr-doped CsFe$_2$As$_2$ measured without the applied magnetic field. (b) Temperature-dependent in-plane *dc* magnetic susceptibility $\chi_{dc}$ of pristine and Co/Cr-doped CsFe$_2$As$_2$, measured with an applied magnetic field of 0.1 T under both field-cooling (FC) and zero-field cooling (ZFC) modes. The peak temperature $T_g \sim 22$ K is defined as the spin-freezing temperature, below which the FC and ZFC curves begin to bifurcate. (Inset) Enlarged image of low-temperature $\chi_{dc}$ of Cr-doped CsFe$_2$As$_2$.

To further confirm the spin-freezing transition in Cr-doped CsFe$_2$As$_2$, the frequency-dependent *ac* magnetic susceptibility was also examined in this work. As shown in Fig. 3(a), a peaklike behavior with peak temperature $T_f$ is observed on each curve of the real part of *ac* magnetic susceptibility

$\chi'_{ac}(T)$. With decreasing frequency, $\chi'_{ac}(T)$ exhibits an overall increase, and the value of $T_f$ continuously decreases towards $T_g$ (see Sec. S4 of the Supplemental Material [36] for a more quantitative analysis of the frequency-dependent $T_f$). All these results are widely observed in most spin-glass materials [46,47], strongly supporting a spin-freezing transition in Cr-doped CsFe$_2$As$_2$. In addition, the observed magnetic hysteresis loop at 2 K also supports the emergence of a spin-freezing state at low temperatures in Cr-doped CsFe$_2$As$_2$ (see Sec. S5 of the Supplemental Material [36]). Based on the above observations, the insulating-like behavior in Cr-doped CsFe$_2$As$_2$ should originate from the spin-freezing transition, which impedes the emergence of the Fermi-liquid state at low temperatures. As we mentioned before, the emergent Fermi-liquid state in the pristine CsFe$_2$As$_2$ stems from the correlation effects driven by Hund's interaction, which leads to a remarkable enhancement of the effective mass [3,12,14,27]. When the Fermi-liquid state is destroyed by spin freezing, the effective mass should also be reduced. As shown in Fig. 3(b), the low-temperature specific heat ($C_p$) was also measured for Cr-doped CsFe$_2$As$_2$. Compared to the pristine CsFe$_2$As$_2$, the value of the electronic specific heat divided by temperature ($C_P/T$) in Cr-doped CsFe$_2$As$_2$ is largely reduced to ~ 80 mJ/mol · K$^2$ at 2 K (~ 180 mJ/mol · K$^2$ for CsFe$_2$As$_2$ [45]). Moreover, the temperature-dependent difference of $C_p/T$ between Cr-doped and pristine CsFe$_2$As$_2$ also exhibits a peaklike behavior around $T_g$ (see the inset of Fig. 3(b)), indicating that the reduction of $C_p/T$ indeed results from spin freezing. This result strongly supports the competition of the Fermi-liquid state and spin-freezing state in this system.

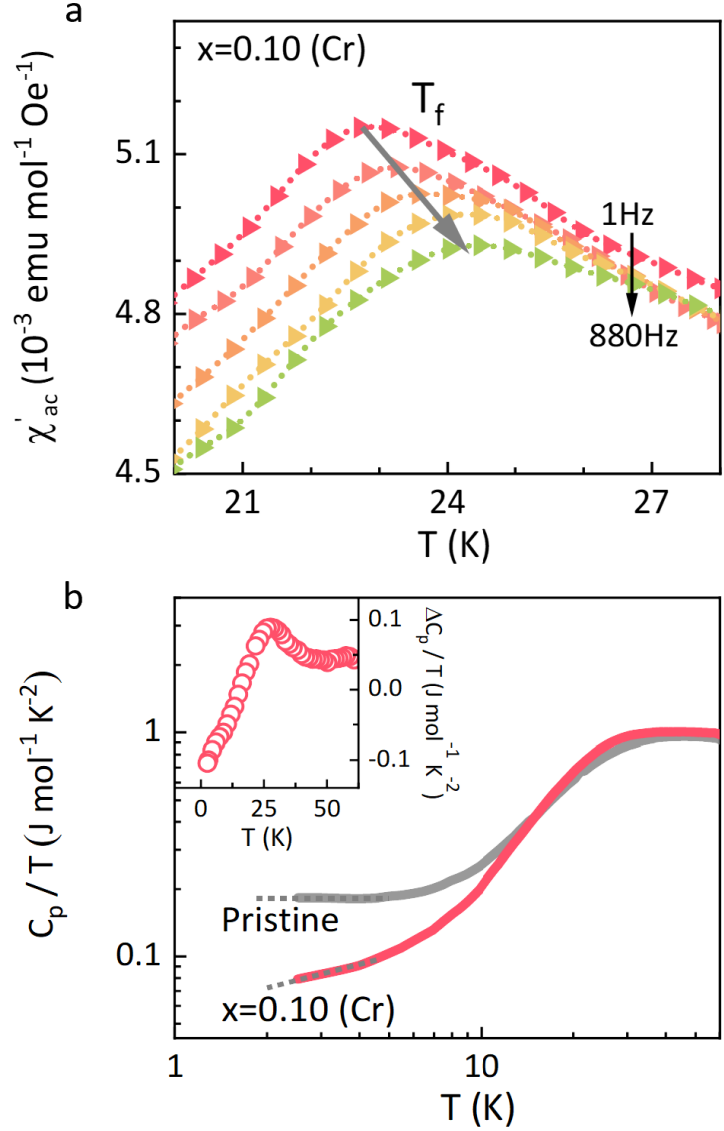

FIG. 3. *ac* magnetic susceptibility and specific-heat evidence for spin freezing and non-Fermi-liquid behavior in Cr-doped $CsFe_2As_2$. (a) Temperature-dependent real part of *ac* magnetic susceptibility $\chi'_{ac}(T)$ for Cr-doped $CsFe_2As_2$. During the measurement at various frequencies (the frequency from the top to bottom curves is 1, 4.5, 120, 620, and 880 Hz, respectively), an *ac* field of 3 Oe is applied. $T_f$ represents the peak temperature, and the gray line just guides the eyes to show the frequency-dependent evolution of $T_f$. (b) Low-temperature specific heat divided by temperature ($C_p/T$) for pristine and Cr-doped $CsFe_2As_2$. (Inset) Temperature-dependent difference $\Delta C_p/T$, which is obtained by subtracting $C_p/T$ of pristine $CsFe_2As_2$ from that of Cr-doped $CsFe_2As_2$ at each temperature.

About the role of Cr doping in $CsFe_2As_2$, the previous studies have already revealed that the Mn/Cr doping can create local moments in $BaFe_2As_2$ system [48], which leads to a change of magnetic structure from stripe-type to G-type antiferromagnetic state [49]. In contrast, the Co/Ni doping can lead to effective electron doping and strongly suppresses the antiferromagnetic state and leads to a superconducting state [50]. In CsFe$_2$As$_2$, we would expect a similar doping effect for Cr and Co doping. In this case, the Cr or Co doping in CsFe$_2$As$_2$ system is expected to create local moments or lead to electron doping, which is qualitatively consistent with the results of magnetic susceptibility in Fig. 2(b). More discussions on this issue will come later.

### B. Nature of spin freezing at the microscale

To further elucidate the nature of spin freezing, a systematic NMR study was performed on both Co-doped and Cr-doped CsFe$_2$As$_2$. Usually, the NMR spectrum represents the distribution of internal fields at nuclei sites [51]. If spin freezing occurs as the temperature decreases, it would lead to a significant broadening of the NMR spectrum [52,53]. As shown in Figs. 4(a) and 4(b), the temperature-dependent NMR spectra on $^{75}$As and $^{133}$Cs nuclei have been measured in Co-doped and Cr-doped CsFe$_2$As$_2$, respectively. Since the internal field produced by magnetic Fe sites leads to a large hyperfine field on $^{75}$As nuclei in the Cr-doped CsFe$_2$As$_2$, the $^{75}$As NMR spectrum would suffer a serious wipe-out effect at low temperatures (see Sec. S6 of the Supplemental Material [36]). For this reason, we chose $^{133}$Cs nuclei with much less hyperfine coupling to Fe sites to study the temperature-dependent evolution of spin freezing in Cr-doped CsFe$_2$As$_2$. As shown in Fig. 5(a), while the NMR linewidth of Co-doped CsFe$_2$As$_2$ exhibits a weak temperature dependence, a strong temperature dependence of the NMR linewidth is observed in Cr-doped CsFe$_2$As$_2$. Simultaneously, the signal intensity of the $^{133}$Cs NMR spectra also wipes out by a factor of 15 as the temperature approaches $T_g$, as shown in Fig. 5(b). As the temperature decreases far below $T_g$, the signal intensity of $^{133}$Cs nuclei is partially restored down to 2 K. In contrast, the NMR results of Co-doped CsFe$_2$As$_2$ do not show any traces for spin freezing. In addition, the present NMR result also rules out the possibility of spin-glass clusters induced by inhomogeneous Cr doping, suggesting a quite uniform distribution of dopant.

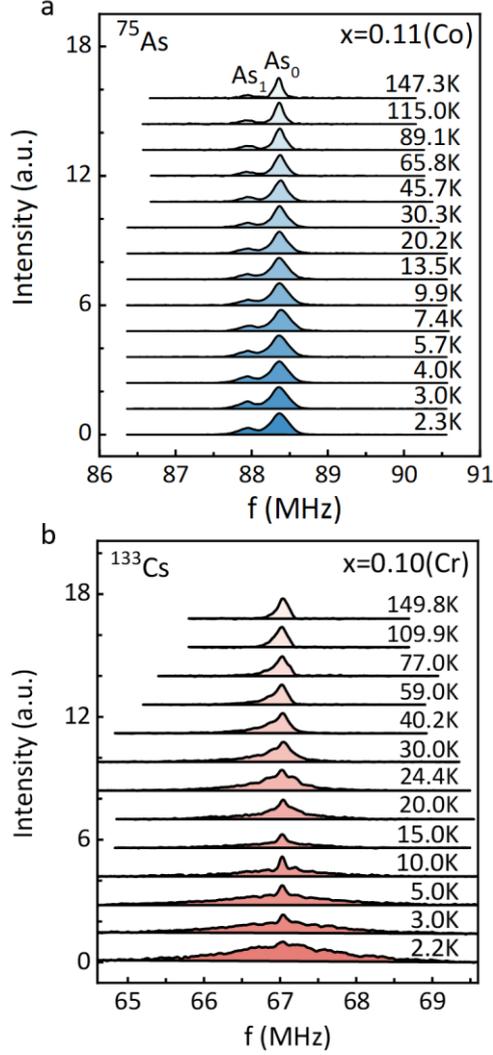

FIG. 4. Temperature-dependent NMR spectra in Co/Cr-doped CsFe$_2$As$_2$. NMR experiments were performed on Co-doped and Cr-doped CsFe$_2$As$_2$ with an applied magnetic field of 12 T parallel to the $ab$ plane. (a) Normalized $^{75}$As NMR spectra for the central transition ($m = +\frac{1}{2} \leftrightarrow m = -\frac{1}{2}$) in Co-doped CsFe$_2$As$_2$. Each spectrum consists of two individual peaks marked by As$_0$ and As$_1$, representing the main As sites with all four nearest-neighbor (NN) sites occupied by Fe atoms and the minor As sites with one of NN sites occupied by Co atoms, respectively [54]. Here, we focus our study mainly on the As$_0$ peak, since the As$_1$ peak is almost temperature independent. (b) Normalized $^{133}$Cs NMR spectra in Cr-doped CsFe$_2$As$_2$. Different from that reported in CsFe$_2$As$_2$ earlier [30], the $^{133}$Cs NMR spectra of Cr-doped CsFe$_2$As$_2$ consist of only one broad peak rather than seven individual peaks even at high temperatures. Considering the small quadrupole frequency $V_Q$ for $^{133}$Cs nuclei, this difference should be caused by a significant magnetic broadening on each individual peak induced by the doping of Cr atoms.

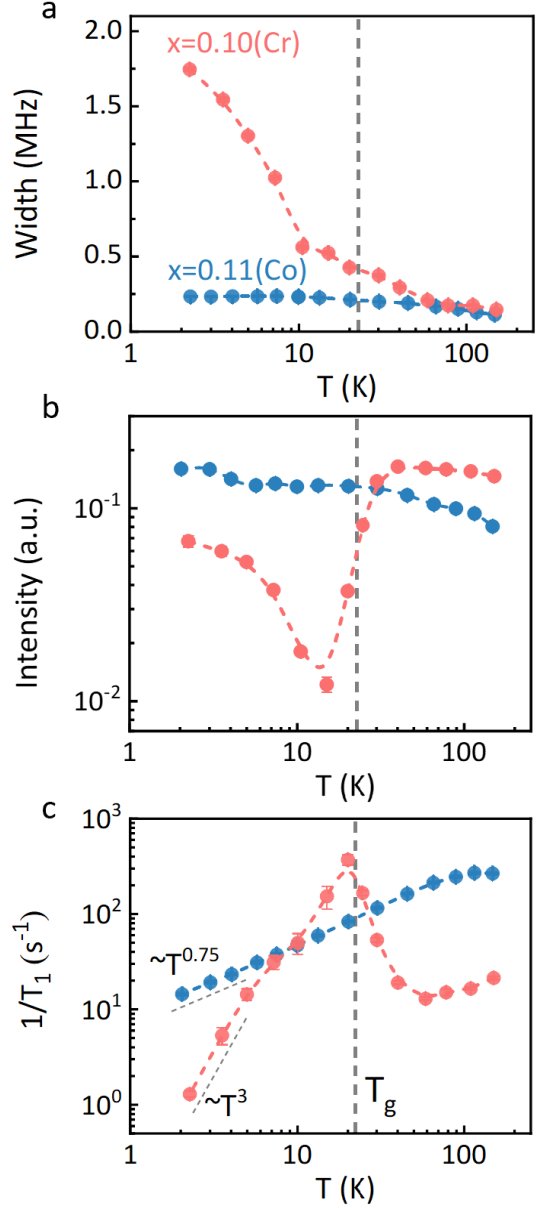

FIG. 5. Temperature-dependent NMR results in Co/Cr-doped $CsFe_2As_2$. (a) Comparison of the temperature-dependent linewidth for the NMR spectra of the $As_0$ peak in Co-doped $CsFe_2As_2$ and for the $^{133}Cs$ NMR spectra in Cr-doped $CsFe_2As_2$. Considering of the different spectral shapes shown in Figs. 4(a) and 4(b), the linewidths of Co-doped and Cr-doped $CsFe_2As_2$ were obtained by Gaussian and Lorentzian fitting, respectively. The error bar from the fitting result is added to each data point of the temperature-dependent linewidth. (b) Comparison of the temperature-dependent signal intensity plus temperature ($I \times T$) normalized by the spin-spin ($T_2$) and spin-lattice relaxation time ($T_1$), where the signal intensity is obtained by integrating the intensity over the full frequency range of the $^{75}As$ NMR spectra for Co-doped $CsFe_2As_2$ and the $^{133}Cs$ NMR spectra for Cr-doped $CsFe_2As_2$. Here, we take the corresponding noise intensity plus temperature as the error bar of the signal intensity plus

temperature. (c) Comparison of the temperature-dependent nuclear spin-lattice relaxation rate $1/T_1$ in Co/Cr-doped $CsFe_2As_2$. For Co-doped $CsFe_2As_2$, the $T_1$ measurement was performed on the $As_0$ peak; for Cr-doped $CsFe_2As_2$, the $T_1$ measurement was performed on the peak frequency of $^{133}Cs$ NMR spectra. The error bar added to each data point of $1/T_1$ is derived from the fitting error of $T_1$ according to the error transfer formula. Here, the vertical gray dotted lines mark the spin-freezing temperature $T_g$, determined from the magnetic susceptibility measurements.

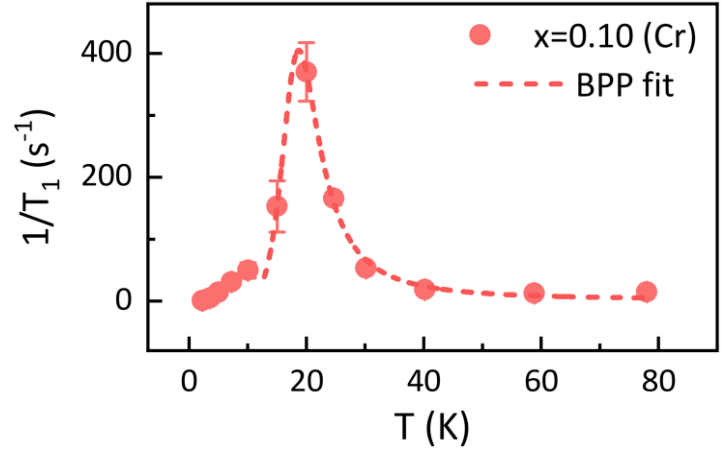

FIG. 6. BPP model fitting of the spin-lattice relaxation rate data in Cr-doped (x=0.10) $CsFe_2As_2$. Temperature-dependent nuclear spin-lattice relaxation rate $1/T_1$ in Cr-doped (x = 0.10) $CsFe_2As_2$. Here, the data of $1/T_1$ are the same as those shown in Fig. 5c, which are plotted on logarithmic axes. Red dashed line is the fit to the BPP model with $\tau_c = \tau_0 \exp(E_a/k_B T)$. Error bar added to each data point of $1/T_1$ is derived from the fitting error of $T_1$ according to the error-transfer formula.

On the other hand, the dynamics of spin freezing could also be examined by measuring the nuclear spin-lattice relaxation. As shown in Fig. 5(c), the nuclear spin-lattice relaxation rate $1/T_1$ of Co-doped $CsFe_2As_2$ shows a characteristic power-law behavior with $1/T_1 \sim T^{0.75}$ below 100 K similar to that of pristine $CsFe_2As_2$ [13], suggesting that the incoherent-to-coherent crossover behavior remains in Co-doped $CsFe_2As_2$. In contrast, the $1/T_1$ of Cr-doped $CsFe_2As_2$ displays an obvious peaklike behavior around $T_g$. Considering the abovementioned spin freezing, the peaklike behavior could be described by the Bloembergen–Purcell–Pound (BPP) model as usually done in other spin-glass systems [51,55]. In the BPP model, the nuclear spin-lattice relaxation rate is expressed as: $\frac{1}{T_1} = \gamma_n^2 \langle h_\perp^2 \rangle \frac{2\tau_c}{1+\omega_n^2 \tau_c^2}$, where $\gamma_n$ is the gyromagnetic ratio, $\tau_c$ is the correlation time, $h_\perp^2 = h_{xx}^2 + h_{yy}^2$ is the transverse hyperfine field, and $\omega_n$ is the NMR frequency. Due to spin freezing, $\tau_c$ should exhibit a divergent behavior as the temperature decreases [52]. In the high-temperature approximation with $\tau_c^{-1} \gg \omega_n$, the above formula of $1/T_1$ can be simplified as $\frac{1}{T_1} = \gamma_n^2 \langle h_\perp^2 \rangle 2\tau_c$. This explains the increasing behavior of $^{133}1/T_1$ above the peak temperature. When $\tau_c^{-1} = \omega_n$, the $1/T_1$ reaches its maximum value and the peak temperature is usually quite close to $T_g$. Upon further cooling, the divergent $\tau_c$ would lead to a decline in $1/T_1$. Here, we fit the BPP model to the data of $^{133}1/T_1$ in the Cr-doped (x = 0.10) CsFe$_2$As$_2$ with an activated correlation time $\tau_c = \tau_0 \exp(E_a/k_B T)$, where $\tau_0$ is the intrinsic correlation time, $k_B$ is the Boltzmann constant and $E_a$ is the effective "energy barrier" for the activation of a thermally driven spin-freezing process. Fitting the data for $T_g < T < T_{onset}$ as usually done in the spin-glass systems [52], we extracted $\tau_0 \sim 20$ μs and $E_a \sim 169$ K for the Cr-doped (x = 0.10) CsFe$_2$As$_2$. Here, the characteristic temperature $T_{onset}$ is considered as the onset temperature of the spin-glass state, below which $1/T_1$ starts to increase with decreasing temperature. Besides, the fitted temperature at which $1/T_1$ is at its maximum value coincides with the experimental data, as shown in Fig. 6. And, the observed $^{133}1/T_1$ indeed reaches its maximum around $T_g$ and then drops rapidly by following a power-law behavior with $1/T_1 \sim T^3$, as shown in Fig. 5c. Such results support that our data of $1/T_1$ could be described well by the BPP model in the Cr-doped (x = 0.10) CsFe$_2$As$_2$. In addition, the temperature-dependent stretching exponent of $T_1$ indicates an inhomogeneity appearing together with spin freezing in Cr-doped CsFe$_2$As$_2$ (see Sec. S7 of the Supplemental Material [36]). All these results are the fingerprints for spin freezing, which have been widely observed in spin-glass materials [52,53,55].

### C. Response of charge degrees of freedom to spin freezing

To clarify the effect of spin freezing on charge degrees of freedom, we have also performed the ARPES experiments on three different samples. As shown in Figs. S9(a) and S9(b), the Fermi surface mappings recorded on Co-doped and Cr-doped CsFe$_2$As$_2$ at 15 K are consistent with earlier reports of K/Rb/CsFe$_2$As$_2$ [56-59]. Here, we focus on the hole pocket (the $\alpha$ band) around the $\Gamma$ point, which shows high intensity around the Fermi level ($E_F$) with nearly degenerate $d_{xz}/d_{yz}$ characteristics. The ARPES intensity plots along the high-symmetry directions of $\Gamma - M$ and $\Gamma - X$ were also recorded on different samples (for details, see Fig. S10 in the Supplemental Material [36]). In the pristine and Co-doped CsFe$_2$As$_2$, band dispersions are almost the same along the two high-symmetry directions. Then, as shown in Figs. 7(a)-7(f), the ARPES intensity plots taken from Cr-doped, pristine, and Co-doped CsFe$_2$As$_2$ at 15 K and 70 K could serve as references for band dispersions along $\Gamma - X$. More clearly, the corresponding band dispersions are plotted in Figs. 7(g) and 7(h). By comparison, the band dispersion of Co-doped CsFe$_2$As$_2$ is almost the same as that of the pristine CsFe$_2$As$_2$ with the Fermi wave vector $k_F \sim 0.27$ Å$^{-1}$ at 15 K and $k_F \sim 0.24$ Å$^{-1}$ at 70 K. This result denotes that the size of the hole pockets for the $\alpha$ band increases at low temperature in Co-doped and pristine CsFe$_2$As$_2$. Similar temperature-dependent shifts for the $\alpha$ band have also been reported in KFe$_2$As$_2$ from previous ARPES measurements [59]. In contrast, there is almost no change in the band dispersions of Cr-doped CsFe$_2$As$_2$ with decreasing temperature, and the Fermi wave vector ($k_F \sim 0.2$ Å$^{-1}$) is smaller than that of pristine CsFe$_2$As$_2$ (Figs. 7(g) and 7(h)). Such a decrease in $k_F$ with the nominally hole doping in Cr-doped CsFe$_2$As$_2$ is beyond a rigid-band model [60], which will be discussed below.

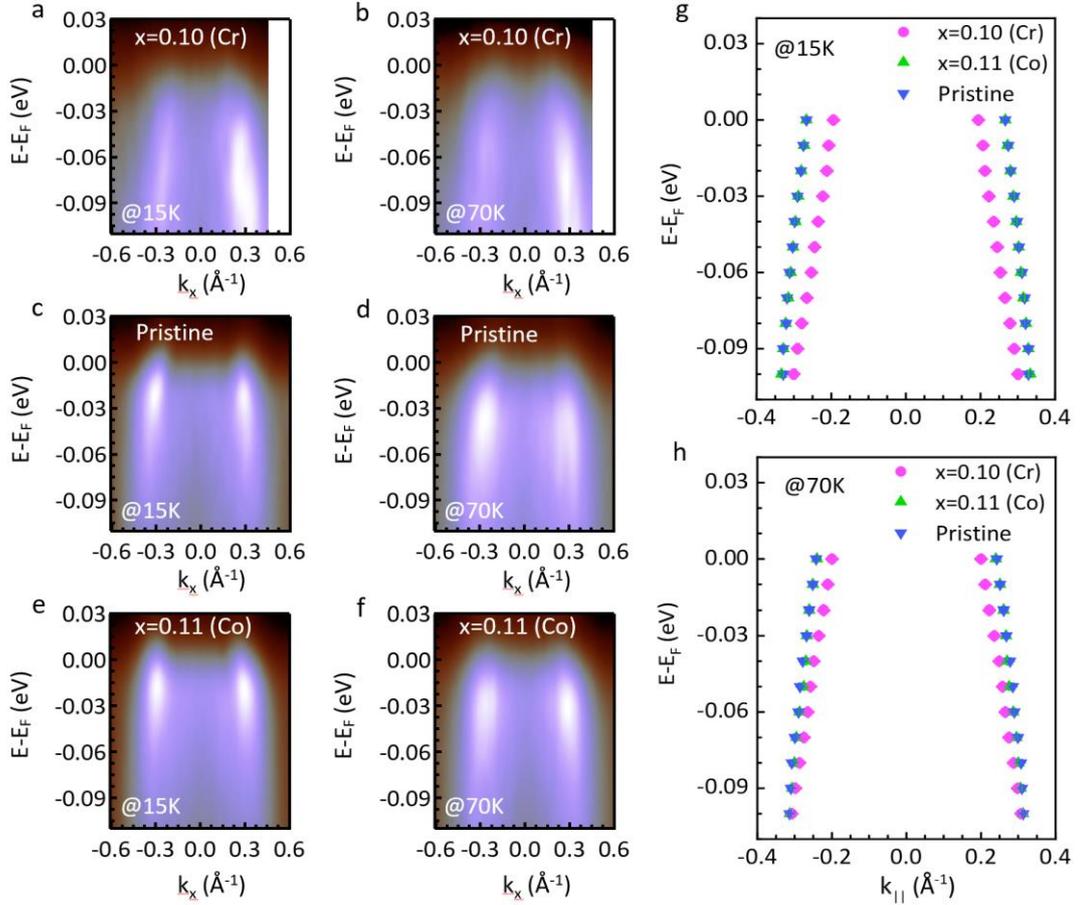

FIG. 7. Comparison of ARPES data at 15 K and 70 K in pristine and Co/Cr-doped CsFe$_2$As$_2$. (a)-(f) ARPES intensity plots of Cr-doped CsFe$_2$As$_2$ (along $\Gamma - X$) and pristine and Co-doped CsFe$_2$As$_2$ (along $\Gamma - M$) as a function of wave vector and binding energy. (g), (h) Comparison of band dispersions from different samples taken at 15 K and 70 K respectively, extracted from peak positions on each momentum distribution curve of the ARPES intensity plots shown in panels (a)-(f). Here, error bars are also plotted, which are smaller than the size of the symbols in panels.

## IV. DISCUSSIONS AND CONLUSIONS

In AFe$_2$As$_2$ (A = K, Rb, Cs), the electronic correlation exhibits strong orbital differentiation, and the strongest electronic correlation is located on the $d_{xy}$ orbital [3,12,21]. At high temperatures, frozen moments should be valid to describe the spin degrees of freedom in the incoherent state, which is quite consistent with Curie-Weiss-like behavior in spin susceptibility [5,11,12,13,25] (see Sec. S9 of the Supplemental Material [36]). As the temperature decreases, the screening of frozen moments due to the coupling of spin and orbital degrees of freedom has been proposed to account for the emergent Fermi-liquid state at low temperatures [8,11,22]. In this picture, the increase in $k_F$ as the temperature decreases in pristine and Co-doped CsFe$_2$As$_2$ should be the consequence of the screening of frozen moments, similar to the Kondo screening in heavy fermion materials [31,61]. In fact, as previously reported in CsFe$_2$As$_2$ [58,59], the more localized $d_{xy}$ orbital (mainly for the $\beta$ band around the $\Gamma$ point) is barely visible due to its weak intensity. However, as reported earlier [12,15,16,22], the more itinerant $d_{xz}/d_{yz}$ orbital would be strongly hybridized with the localized $d_{xy}$ orbital at low temperatures in iron-based superconductors, since they all cross the Fermi energy and are the main contributions for the Fermi surfaces. Therefore, it is reasonable to conjecture that, the $\alpha$ band with $d_{xz}/d_{yz}$ character could be affected by the $\beta$ band with $d_{xy}$ character. This explains the temperature-dependent evolution of the $\alpha$ band during the screening of frozen moments in the pristine and Co-doped CsFe$_2$As$_2$. Considering the spin-freezing transition around $T_g$, the doping of Cr atoms should enhance the coupling of frozen moments, which is a competing interaction for the abovementioned screening interaction [11,31]. This effect could contribute to the temperature-independent $k_F$ in the Cr-doped CsFe$_2$As$_2$. Meanwhile, the emergence of a spin-freezing state at low temperatures indicates that the electronic state is incoherent in Cr-doped CsFe$_2$As$_2$, explaining why $k_F$ for the $\alpha$ band is smaller in Cr-doped CsFe$_2$As$_2$ compared to that of pristine and Co-doped CsFe$_2$As$_2$. However, this picture still needs more studies from alternative perspectives.

On the other hand, the insulating-like behavior of resistivity in Cr-doped CsFe$_2$As$_2$ suggests a non-Fermi-liquid ground state. To our knowledge, the low-temperature upturn in resistivity could originate from several quantum confinement effects, such as weak localization, the Kondo effect or electron-electron interactions [62,63]. Nevertheless, these mechanisms could be ruled out in our case by magnetoresistance measurements (see Sec. S10 of the Supplemental Material [36]). Here, the persistence of the wipe-out effect in the NMR spectrum at 2 K suggests that such an insulating-like behavior in Cr-doped $CsFe_2As_2$ could still be affected by the scattering from spin fluctuations in the spin-freezing state, as described in other spin-glass systems [64]. Finally, the present study unambiguously confirms the competition between the Fermi-liquid state and magnetic ordering in the case of Hund's metal, which deserves more investigations in the future.

ACKNOWLEDGMENTS

This work is supported by the National Key R&D Program of the MOST of China (Grant No. 2022YFA1602601), the National Natural Science Foundation of China (Grants No. 11888101, No. 12034004, No. U2032153), the Strategic Priority Research Program of Chinese Academy of Sciences (Grant No. XDB25000000), the Anhui Initiative in Quantum Information Technologies (Grant No. AHY160000), the Innovation Program for Quantum Science and Technology (Grant No. 2021ZD0302800)，the CAS Project for Young Scientists in Basic Research (2022YSBR-048).